\documentclass[aip,pop,reprint,numerical,twocolumn,superscriptaddress]{revtex4}
\usepackage{amssymb, amsmath,framed}

\usepackage{bm}
\usepackage[colorlinks=true,linkcolor=blue]{hyperref}
\expandafter\ifx\csname package@font\endcsname\relax\else
 \expandafter\expandafter
 \expandafter\usepackage
 \expandafter\expandafter
 \expandafter{\csname package@font\endcsname}
\fi
\def\bq{\begin{equation}}
\def\eq{\end{equation}}
\def\bqy{\begin{eqnarray}}
\def\eqy{\end{eqnarray}}

\def\be{\beta}

\def\de{\delta}

\def\ep{\epsilon}

\def\ka{\kappa}

\def\na{\nabla}

\def\Om{\Omega}
\def\p{\partial}

\def\rh{\rho}

\def\p{\partial}

\def\R{\mathbb{R}}

\def\calf{\mathcal{F}}

\def\calj{\mathcal{J}}
\def\calk{\mathcal{K}}
\def\call{\mathcal{L}}

\def\calo{\mathcal{O}}

 \def\q#1#2{q^{\hspace{1pt}#1}_{\,,\hspace{1pt}#2}}
 \def\a#1#2{a^{\hspace{1pt}#1}_{\,,\hspace{1pt}#2}}
 \def\d#1#2{\delta^{\hspace{1pt}#1}_{\,,\hspace{1pt}#2}}

\begin{document}

\title{Hamiltonian and action formalisms for two-dimensional gyroviscous MHD}

\author{P.~J.~Morrison}
\email{morrison@physics.utexas.edu}
\affiliation{Department of Physics and Institute for Fusion Studies, The University of Texas at Austin, Austin, TX 78712}
\author{M.~Lingam}
\email{manasvi@physics.utexas.edu}
\affiliation{Department of Physics and Institute for Fusion Studies, The University of Texas at Austin, Austin, TX 78712}
\author{R.~Acevedo}
\email{raul\_ace60@yahoo.com}
\affiliation{3311 Black Locust Dr., Sugar Land, Texas 77479}
 
\date{March 2014}
\begin{abstract}
A general procedure for constructing action principles for continuum models via  a generalization of  Hamilton's principle of mechanics is described. Through the procedure, an action principle for a gyroviscous magnetohydrodynamics (MHD) model is constructed.  The model is shown to agree with a reduced version of Braginskii's fluid equations. The construction reveals the origin of the gyromap, a device  used to derive previous gyrofluid models.  Also, a systematic reduction procedure is presented for obtaining the  Hamiltonian structure in terms of the noncanonical Poisson bracket.  The construction procedure yields a class of Casimir invariants, which are then used to variational principles for equilibrium equations with flow and gyroviscosity.   The procedure for obtaining reduced fluid models with gyroviscosity is also described. 
\end{abstract}

\maketitle


\section{Introduction}
\label{sec:intro}

The main purpose of this paper is to describe a very general procedure for constructing action principles for plasma models, and then to use the procedure  to  construct a gyroviscous magnetofluid model,   a version of which agrees with a two-dimensional reduced ideal limit of Braginskii's equations \cite{braginskii65}.   The action principle leads naturally to an unambiguous conserved energy functional,  with the associated  Hamiltonian description.  In addition,  a family of constants of motion is also obtained, which can be used for construction of  variational principles for equilibria and $\de W$ type stability criteria.   Another by-product  obtained  is the derivation and physical identification of the gyromap, a tool introduced in \cite{MCT84} and used in previous derivations of reduced fluid models  \cite{HHM86,HHM87,ICTC11}.  It is shown how to directly obtain such reduced fluid models from the action principle. 

The construction and use of Hamiltonian and action principle (HAP) formulations of continuum models possess a fascinating   history that dates back to Langrange's pioneering work in analytical mechanics \cite{Lagrange},   which   was extended by many  illustrious scientists  (e.g., \cite{clebsch57,clebsch59,helmholtz58,hankel,kirchhoff76})  in the 19th century. In the 20th century,  important  seminal contributions,  too numerous to list (e.g., \cite{serrin59,newcomb62,newcomb73}), were obtained, and in 1980s renewed interest in HAP formulations  arose from the key work \cite{morrison80}. Reviews of this approach can be found in \cite{morrison82,morrison98,morrison05,morrison06}. 

Evidently, the HAP formalism has a distinguished history --  it  is also  the conventional basis for building models in theoretical physics and describing how to do  this in a plasma physics context  is a major goal of the present work.    There  are many reasons for constructing  models via an action principle, apart from the aesthetics  and simplicity afforded by the approach.   Since  the underlying basic physics possesses a HAP formalism,  viz.\  the relativistic $2N$-body problem of $N$ electrons and $N$ ions interacting via  the core electromagnetic interaction,  one expects the nondissipative versions of simplified models to inherit  this underlying HAP structure.  This is the case for the most important equations of plasma physics, e.g.,   magnetohydrodynamics (MHD) \cite{morrison80}, the Vlasov description (e.g., \cite{morrison82,morrison13}), and the BBGKY hierarchy \cite{MMW84}  all possess HAP structure.    Another reason is that  building an action principle, by the steps we describe here, is easier than the usual perturbative or alternative model building approaches.   And, an  oft stated advantage is the clear emergence of energy and other invariants of motion for  the nondissipative dynamics, which will be exemplified here by the discovery of new  such invariants.

The elimination of unphysical dissipation is an important advantage of  the HAP formalism over other approaches. It is common to start with a parent model, and perform an ordering in the equations of motion, to obtain a ``reduced" model. However, in this process of obtaining simplified models, one   runs the risk of introducing ``false" dissipation (e.g., \cite{HKM85,HKM86, tronci, kimura} for such cases); i.e.,  the feature of energy conservation can be  broken through an improper phenomenological prescription. By using the HAP formalism, an ordering can be done  in a  rigorous manner that  ensures  the resultant model is non-dissipative, when it should be.

The process of obtaining reduced fluid models lends itself well to the  HAP approach.  These kinds of models are two or nearly two-dimensional magnetofluid models that have been developed to incorporate important physics into computable closed dynamical systems.  One of the early examples   is reduced MHD \cite{strauss76}, but many models for many purposes have been obtained over the years (e.g., 
\cite{MH84,MM84,HKM85,HHM87,KPS94,KK04,TMG07,TMWG08,WMH04,WHM09}).  In the past, reduced fluid models mostly have  been obtained through an asymptotic expansion, through suitable orderings, or through ad hoc approaches.     (See \cite{TM00,tassi10} for general theory of models of this type.)\  \ In this paper, we will show how the HAP formalism presents a clear-cut manner to obtain reduced fluid models  with gyroviscous effects.  This is done in the context of a three field high-$\beta$ reduced MHD (RMHD) model.  Also, as noted above, we will explain the origin of the gyromap, a tool introduced in \cite{MCT84} and used in previous derivations of reduced fluid models  \cite{HHM87,ICTC11}

The remainder of the paper is organized as follows.  In Sec.~\ref{hamsPrinc},  we commence with a description of the  action principle suitable for the derivation of  our continuum models of interest here.  In Sec.~\ref{LagEul},  the Lagrangian and Eulerian descriptions of a fluid are reviewed, and the rationale behind choosing the former as a starting point  is presented. In Sec.~\ref{buildinggeneral},  we describe the  general procedure for building action principles.  In Sec.~\ref{building},  we describe how a gyroviscous model is constructed by using   this  approach,   present the resultant equations of motion, and compare them to those of Braginskii \cite{braginskii65}.  In Sec.~\ref{HamDescgen}, we describe the transformation from the action principle description of Sec.~\ref{building} to the Eulerian variable Hamiltonian description in terms of noncanonical Poisson brackets. In Sec.~\ref{Casimirs}, we present  Casimirs for the full and reduced  models, and describe  equilibria that are obtained via the energy-Casimir method.   In this way equilibrium equations with flow and the influence of  gyrovisous effects are obtained. Here we also describe how to derive reduced fluid models.  Finally, we summarize and conclude with some comments on the HAP formalism in Sec.~\ref{Conclusion}.


\section{Hamilton's Principle and the functional derivative}
\label{hamsPrinc}

Most textbooks follow a similar prescription for deploying Hamilton's principle. They  begin by  identifying a configuration space and variables that describe the system in its entirety,  the generalized coordinates  $q^{i}(t)$, 
where $i=1,2,\dots,\, N$ and $N$ is the number of degrees of freedom of the
system.  Then the Lagrangian  $L:=T -V$ is obtained by identifying the kinetic energy  $T$   and potential energy $V$, yielding the  action functional 
 \bq S[{q}] = \int_{t_1}^{t_2}  \!dt \, L(q,\dot q,t)\,.
\label{action}
\eq 
By ``functional", we refer to a quantity whose domain is comprised of functions 
and whose range is given by real numbers.  In other words, for a given path $q(t)$, the action functional $S[{q}]$ returns a 
real number upon substitution of the path into the above expression.

In  Hamilton's principle  the lower and upper limits of the path, $q(t_1)$ and $q(t_2)$, are fixed and   the path that gives rise to the extremal value is sought.  Extremal means that the functional derivative of
the action vanishes, $\de S[q]/\de q^{i} = 0$,
where the functional derivative is defined by
 \bqy
\de S[q;\de q]&=&
\left.\frac{d S[q +\ep \de q]}{d \ep}\right|_{\ep =0}
=: \left\langle \frac{\de S[q]}{\de q^{i}},\de q^{i}\right\rangle
\nonumber\\
&=& \int_{t_1}^{t_2}\!dt \left(\frac{\p L}{\p q^{{i}}} -\frac{d}{d t}\frac{\p L}{\p \dot{q}^{{i}}}\right) \de q^{{i}} 
\,.
\label{functderiv}
\eqy
In the above expression $\de q(t)$ is an arbitrary perturbation of a path $q(t)$; given the arbitrariness,  the only way for $\de S$ to vanish for all $\de q(t)$ is to have the quantity 
within the parentheses vanish, i.e. 
 \bq
\frac{\de S[q]}{\de q^{i}} = 0\quad \Leftrightarrow\quad 
\frac{\p L}{\p q^{i}} -{d\over dt}{\p  L\over \p 
\dot q^{i}}=0\,.
\eq
In other words,   the extremal path corresponds to the  Euler-Lagrange equations of motion.

 \section{Lagrangian and Eulerian descriptions - attributes, observables and the  Lagrange to Euler map}
 \label{LagEul}

In this section we review the Lagrangian and Eulerian descriptions of a fluid and their relationship. 
The section is divided into two  parts. In Sec.~\ref{ssec:lagVar} we describe the basic Lagrangian variable that describes a  trajectory of a fluid element, and then present some useful algebraic identities and properties.  This description most naturally possess  HAP formulations, since it effectively treats the fluid as a set of particles.  Then, in Sec.~\ref{ssec:AOLE},  we explore  the relationships between the intrinsic properties of a fluid element and their Eulerian  observable counterparts via the Lagrange to Euler map  that relates the two descriptions.  For background material we suggest \cite{serrin59,newcomb62,VKF67,morrison98}.

\subsection{The Lagrangian variable $q$ and its properties}
\label{ssec:lagVar}

The {\it Lagrangian variable} can be understood as a coordinate that denotes the position of a fluid element or parcel,  as it is sometimes called,   at a given time $t$.  The coordinate, which indicates the position relative to some origin is denoted  by  $q=q(a,t)=(q^1,q^2,q^3)$; for the sake of simplicity Cartesian coordinates are used.  The quantity  $a=(a^1,a^2,a^3)$ denotes the {\it fluid element label} at time $t=0$, which implies that $a=q(a,0)$, but this means  of labeling need not always be the case (cf.\ \cite{amp1}).  In general, the continuous label $a$ is analogous to the discrete index that enables us to keep track of a given particle in a finite degree-of-freedom system.   Suppose  $D$ denotes  the domain that is fully occupied by the fluid, then the map  $q\colon D\rightarrow D$  is assumed to be 1-1 and onto at a given fixed time $t$.  We will suppose that $q$ is invertible and smooth and any other ``nice" properties that the problem demands.

Given the Lagrangian coordinate $q$, we introduce two other related important quantities: the deformation matrix, $\p q^i/\p a^j=:\q ij$ and the corresponding determinant, the Jacobian, $\mathcal{J}:= \det(\q ij)$,  which in three and two dimensions,  respectively,  is given by
\bqy
\mathcal{J}&=& \frac1{6}\ep_{kjl}\ep^{imn}    \q ki \q jm \q ln\,,
\label{J3}
\\
&=& \frac1{2}\ep_{kj}\ep^{il}  \q ki \q jl \,,
\label{J2}
\eqy
where  $\ep_{ijk}=\ep^{ijk}$ and $\ep_{ij}=\ep^{ij}$ are  the Levi-Civita tensors in the appropriate number of dimensions.
Assuming  the label specifies  a unique trajectory, we conclude that  $\mathcal{J}\neq 0$;  this ensures the invertibility of $q=q(a,t)$, denoted by $a=a(q,t)$. The quantity $a(q,t)$ can be understood to be the label of a fluid  element that reaches $q$ at time $t$.  In general for  coordinate transformations we have 
\[
\q ik\a kj =\a ik\q kj =\d ij\,,
\]
i.e.\  the the deformation matrix has an inverse given by  $\a kj= \p a^k/\p q^j$ where repeated indices are summed.   Using  $q(a,t)$ or  its inverse, we can express quantities such as $\a kj$ as functions of either $q$ or $a$.

A volume element $d^3a$ at  time $t=0$  maps into the volume element at time $t$ according to  
\bq
d^3q=\mathcal{J} d^3a\,,
\label{vol}
\eq
and the components of an  area element evolve according to  
\bq
(d^2q)_i=  \mathcal{J} \a ji \, (d^2a)_j\,,
\label{area}
\eq 
where $\mathcal{J} \a ji$ is the transpose of the cofactor matrix of $\q ji$ given by
\bq
\mathcal{J} \a ik={1\over 2}\ep_{kjl}\ep^{imn}\q jm   \q ln  
\quad\mathrm{or}\quad
\mathcal{J} \a ik =\ep_{kj}\ep^{il} \q jl \,,
\label{co2}
\eq
 in three and two dimensions, respectively.
Other useful identities are 
\bqy
\frac1{\calj}\frac{\p \mathcal{J}}{\p \q ij }&=& \a ji  \,,
\label{dedet} \\
\frac{\p (\mathcal{J} a^i_{\, , k})}{\p a^{i}}&=& 0\,,
\label{detco}
\eqy
where (\ref{dedet}), the standard rule for differentiation of determinants,  follows from (\ref{J3})  or (\ref{J2}), and  (\ref{detco}) follows from   (\ref{co2}) by the antisymmetry of $\ep_{ijk}$ or $\ep_{ij}$.


\subsection{Attributes, observables,  and the Lagrange to Euler map} 
\label{ssec:AOLE}

Up to now, we have  considered  kinematical properties of the fluid, as described by the Lagrangian coordinate $q$. But, a fluid element is not solely characterized by its position $q$ and its label $a$. In addition, it is  endowed with  certain intrinsic properties, such as its density or, in the case of MHD,  some  magnetic flux it might carry unchanged.  Thus, it is natural to investigate  a general characterization of such intrinsic properties.   Here we do this for the case of three spatial dimensions. 

We will refer to quantities that the fluid element carries as  {\it attributes},  since they are intrinsic to the fluid under consideration. A fluid element that starts off at time $t=0$ carries its  attributes unchanged.  Thus,  by definition attributes are purely functions of the label $a$,  and are Lagrangian variable constants of motion. We will use the subscript `$0$' to distinguish attributes from their Eulerian counterparts, discussed below.

Usually in fluid mechanics  the Lagrangian variable description is not emphasized and, consequently,  attributes are usually not discussed. More typically,  it is the Eulerian fields that are emphasized and observed. We will refer to such as Eulerian observables, or just  {\it observables}  for short.  The observables, being  Eulerian,  vary in space and time, and are therefore functions of $r:=(x,y,z)=(x^1,x^2,x^3)$ and $t$.  
   
Some of the most commonly used  Eulerian variables include velocity field $v(r,t)$ and the mass density $\rho(r,t)$.   We reiterate that it is crucial to distinguish  the Lagrangian coordinate $q$ from the  Eulerian observation point  $r$. The latter is an independent variable that does not move with the fluid, although it is  a point  of $D$.  The inability or unwillingness to distinguish between the two descriptions has led to confusion in the literature.  Therefore, it is important  to ask,   how precisely are  the two descriptions  related to one another?
 
Given knowledge of $q(a,t)$,  the observables are  uniquely determined. The rules for  this determination are based on the nature of the attributes, in particular,  their tensorial properties.    For example, consider the velocity field $v(r,t)$.  If we were to insert a velocity probe into a fluid at $(r,t)$, we would measure the velocity of the fluid element that happened to be at  $r$ at time $t$.  Hence,   $\dot{q}(a,t)=v(r,t)$, where the overdot indicates that the time derivative is obtained at fixed $a$.  We are still left with the ambiguity of determining the label $a$, but the element at $r$ is given by $r=q(a,t)$, whence  $a=q^{-1}(r,t)=:a(r,t)$.  By combining all this information, we see that the Eulerian velocity field is given by
\bq
v(r,t) =\left.\dot{q}(a,t)\right|_{a=a(r,t)}\,.
\eq
The above expression is an example of the Lagrange to Euler map that supplies  a means of moving from one picture to the other.

Attributes, as part of their definition, possess rules for transformation to their corresponding Eulerian observable.  The totality of these rules determines the set of observables.  For a continuum system, in which mass is neither created nor destroyed,  it is natural to attach a mass density, $\rho_0(a)$, to the element labeled by $a$.  We note that the mass in a given volume is given by $\rho_0d^3a$. By demanding that the mass by conserved, regardless of whether one uses the Eulerian or Lagrangian picture, we see that \ $\rho(r,t)d^3r=\rho_0d^3a$.  By using (\ref{vol}) we obtain $\rho_0=\rho \calj$.
This defines the rule for transforming to the Eulerian description, which here amounts to $\rho$ defining a three-form.
Similarly, we may attach a magnetic field $B_0(a)$ to a given fluid element, and define its transformation law by insisting on frozen in flux. This yields $B\cdot d^2r=B_0\cdot d^2a$, and from (\ref{area}) we obtain $\calj B^i=\q ij  \,B_0^j$.  This condition amounts to $B$ defining a two-form.  Evaluating the expressions for the mass density and the magnetic field at $a=q^{-1}(r,t)=:a(r,t)$, then gives  the  {\it Lagrange to Euler map} for these quantities. In other words,  given $q(a,t)$ and the attributes,  the fields $\{\rho,v,B\}$, the observables,  are now  defined.  

Most of the time  we will find it convenient to  work with the alternative set of observables $\{\rho, M,B\}$, where $M=\rho v$ is the momentum density.  This allows a  convenient way to represent the Lagrange to Euler map  in terms of  integrals of over a Dirac delta function, which  is used as a probe to `pluck out' the fluid element that happens to be at the Eulerian observation point $r$ at time $t$.  As an example of this procedure, the mass density $\rh(r,t)$ is obtained by
\bqy
\rh({r},t)&=&\int_D \!d^3a
\, \rh_0(a) \, \de\left({r}-{q}\left(a,t\right)\right)
\nonumber\\
&=&\left. \frac{\rh_0}{\mathcal{J}}\right|_{a=a({r},t)}\,.
\label{rhoEu3D}
\eqy
We will introduce the  momentum density, $M^c=(M^c_1,M^c_2,M^c_3)$, which is related to the Lagrangian canonical momentum through the expression
\bqy
M^c(r,t)&=&\int_D \!d^3a \,
{\Pi}(a,t) \, \de\left({r}-{q}(a,t)\right) 
\nonumber\\
&=& \left.
\frac{\Pi(a,t)}{\mathcal{J}}
\right|_{a=a(r,t)}\,.
\label{Mcan3D}
\eqy
The superscript `$c$' indicates that the momentum density constructed is the canonical one, as opposed to a different momentum density introduced in the next section. For   MHD,     $\Pi(a,t)=(\Pi_1,\Pi_2,\Pi_3)=\rh_0 \dot{q}$.  In general, note that $\Pi(a,t)$ can be found from the Lagrangian through $\Pi(a,t) = {\de L}/{\de \dot{q} }$ and is not always equal to  $\rh_0 \dot{q}$. Lastly, 
\bqy
B^i(r,t) &=& \int_D \!d^3a \, \q ij(a,t) B_0^j(a) \, 
\de\left({r}-{q}(a,t)\right) \nonumber  \\ 
&=& \left.
\q ij(a,t)\frac{B_0^j(a)}{\mathcal{J}} 
\right|_{a=a(r,t)}\,,
\label{Brel}
\eqy
for the components of the magnetic field.

We  round off this subsection with a mention of a few additional useful identities that  play a role in the subsequent sections.  The chain rule reveals  the components of  the 
Eulerian gradient are given by
\bq
\frac{\p}{\p x^k}=\left. \a ik  \, \frac{\p}{\p a^i}\,\right|_{a=a(r,t)}\,.
\label{ELgrad}
\eq
With the condition that $r=q(a,t)$, the time derivative of any function $f(a,t)=\tilde{f}(r,t)= \tilde{f}(q(a,t),t)$ 
can be mapped to the corresponding Eulerian variables according to the expression
\bqy
 \left.\dot f \, \right|_{a=a(r,t)}&=&  \frac{\p \tilde{f}}{\p t} + \left.  \dot{q}^i(a,t)\, \frac{\p \tilde{f}}{\p x^i}\,\right|_{a=a(r,t)}
 \nonumber\\
 &=& 
 \frac{\p \tilde{f}}{\p t} + v\cdot \nabla   \tilde{f}(r,t)\,.
 \label{lagDet}
\eqy
As stated earlier, we note that the overdot denotes the  time derivative at constant $a$,     $\p/\p t$   denotes the  time derivative at constant $r$, and $\nabla$ is the Eulerian derivative, i.e.\  $\p/\p r$ with components $\p/\p x^i$ .  

Lastly, we can obtain an evolution equation for the determinant $\calj$ using Eq.~(\ref{dedet})  
\bq
\dot\calj=\frac{\p \mathcal{J}}{\p \q ij } \,    \dot{q}^{\hspace{1pt}i}_{\,,\hspace{1pt}j}=
\calj \a ji   \,  \dot{q}^{\hspace{1pt}i}_{\,,\hspace{1pt}j}\,,
\eq
which upon evaluation at $a=a(r,t)$ gives  a formula due 
to Euler \cite{serrin59}, 
\bq
\frac{\p \tilde\calj}{\p t} + v\cdot \nabla \tilde \calj= \tilde\calj \, \na\cdot v\,.
\label{Jeuler}
\eq


\section{A general procedure for building an action principle for continuum models}
\label{buildinggeneral}

In this section, we provide a brief summary of the general methodology advocated in  \cite{morrison09} for  building action principles for continua. One of the major advantages of building an action principle from scratch deserves a mention before proceeding further. As opposed to ordering in the equations of motion or {\it ad hoc} methods that are deployed in obtaining models from a basic set of equations, we  proceed to introduce each term in the action serially, and emphasize the physical relevance of each to the model being built. This allows for  improved physical understanding and motivation as to why the different terms arise, and what roles they play in the model.

The first step in constructing an action principle lies in choosing the domain $D$.  For a fluid it would be either one, two, or three-dimensional, $D\subset \R^{1,2,3}$.  Furthermore, we  suppose that there exists a Lagrangian (trajectory) variable $q\colon D\rightarrow D$.  We also suppose that $q(a,t)$, where the label $a\in D$, is a well behaved function that is smooth, has an inverse, etc. 

The next step lies in choosing the sets of attributes and the corresponding observables, defined via a  Lagrange to Euler map. There is some  freedom in choosing the set of observables that interest us, as discussed in the previous section. It is  important to recognize that the observables must be completely determined by the functions $q(a,t)$, but the converse statement is not a necessity.

From the analogy with Hamilton's action principle in mechanics, it is evident that the action will comprise of terms that   involve the variable $q(a,t)$ and its derivatives  with respect to both its arguments.    The last step of the method is to  impose a most stringent requirement upon the terms in the action --  viz.\  the existence a {\it closure principle} which ultimately means that our theory must be `Eulerianizable.'\ \    More precisely,  we impose the condition that  our action  be expressible entirely in terms of our set of observables. Such a requirement is well motivated, since it leads to energy-like quantities that are  entirely expressible in terms of the desired Eulerian variables.     As an example, note that the kinetic energy functional for a fluid   satisfies 
\bq
T[q]:=\frac1{2} \int_D \!d^3a\, \rho_0(a) |\dot q|^2=
\frac1{2} \int_D \!d^3r\, \rho  |v|^2\,, 
\label{Tq}
\eq
where $|\dot q|^2:=\dot{q}^i g_{ij}\dot{q}^j=\dot{q}^i\dot{q}_i$ and  for cartesian coordinates the metric $g_{ij}=\de_{ij}$ is chosen.   Thus the Lagrangian variable description of the first equality can be written as the purely Eulerian description of the second.  The imposition of the closure principle leads to important consequences:   equations of motion that are purely expressible in terms of our observables, i.e.\ an Eulerian variable description,  and  an Eulerian Hamiltonian description in terms of  noncanonical Poisson brackets, which are discussed in the subsequent sections.


\section{Building an action principle for the 2D gyroviscous  fluid model}
\label{building}

Now we following the method described in Sec.~\ref{buildinggeneral}.  First we introduce and motivate the set of observables,  then we describe how  their corresponding attributes are used to construct an action principle. 


\subsection{The observables of the 2D gyroviscous model}
\label{observeGyro}

  For the  first step,  choosing the domain,  we select $D= \R^2$,  with coordinates $(x,y)$, since our theory is two-dimensional.  Next, when defining the set of observables we must select the  momentum density. We can  choose either the canonical momentum defined in (\ref{Mcan3D}) or  the `kinetic' momentum defined by
\bqy
M(r,t)&=&\int_D \!d^2a \,
\rho_0(a) {\dot{q}}(a,t) \, \de\left({r}-{q}(a,t)\right)\nonumber \\ 
&=& \left. \frac{\rh_0 \dot{q}(a,t)}{\mathcal{J}}\right|_{a=a({r},t)}\,.
\label{Mkin}
\eqy
The 2D version of the canonical momentum defined through (\ref{Mcan3D}) is given by
\bqy
M^c(r,t)&=&\int_D \!d^2a \,
{\Pi}(a,t) \, \de\left({r}-{q}(a,t)\right) 
\nonumber\\
&=& \left.
\frac{\Pi(a,t)}{\mathcal{J}}
\right|_{a=a(r,t)}\,,
\label{Mcan}
\eqy
where we  suppose there is no momentum in the $\hat{z}$-direction. 
In general,  the kinetic  and  canonical momenta are not  the same; in fact, for the action we  construct for our   gyroviscous model, their difference defines the gyromap,  a key result of this paper.   When deriving the equations of motion, we  work with  $M$, although we  make use of $M^c$ when developing the Hamiltonian formalism for this model.  
Next consider the  magnetic field, which will   be in our set of observables.  Since $\nabla \cdot B = 0$, we   decompose it  as follows:
\bq
B = B_z(x,y,t)\, \hat{z} + \hat{z}\times \nabla \psi(x,y,t),
\label{BEuler}
\eq
which is a usual decomposition with $\psi$ representing the parallel vector potential.   Following the same line of reasoning of Sec.~\ref{ssec:AOLE}, the associated  attribute takes on the form
\bq
B_0 = B_{0z}(a)\, \hat{z} + \hat{z} \times \nabla_{a} \psi_0(a)\,.
\label{BLagrange}
\eq
with the  correspondence between these attributes and  observables following from  (\ref{Brel}) and  yielding the Lagrange to Euler correspondences  
\bqy
B_z({r},t)&=&\int_D \!d^2a
\, B_{0z}(a) \, \de\left({r}-{q}\left(a,t\right)\right)
\nonumber\\
&=&\left. \frac{B_{0z}}{\mathcal{J}}\right|_{a=a({r},t)}\, ,
\label{Bz}
\\
\psi({r},t) &=& \left. \psi_0\right|_{a=a({r},t)}\,.
\label{Psi}
\eqy
Our last observable is the density,  which is given by the 2D version of (\ref{rhoEu3D})
\bq
\rh({r},t)=\int_D \!d^2a
\, \rh_0(a) \, \de\left({r}-{q}\left(a,t\right)\right)
=\left. \frac{\rh_0}{\mathcal{J}}\right|_{a=a({r},t)}\,.
\label{rhoEu}
\eq
Thus, our set of observables is  $\{\rho, M, B_z, \psi \}$.

Up to now we have not specified anything about the internal energy per unit mass, which in general is $U:=U(\rho,s)$. However, we  restrict ourselves to the barotropic case, i.e., assume a  thermodynamic  energy that is independent of the entropy $s$.     Thus,  the pressure is obtained from $U:=U(\rho)$ via  $P = \rho^2 {dU}/{d\rho}=\kappa\rho^2$, with $\ka$ constant. 
 The Lagrange to Euler map between $P$ and $P_0$ can be determined through the use of (\ref{rhoEu}); it takes on the form
\bq
P = \left. \frac{P_0} {{\mathcal{J}}^2} \right|_{a=a({r},t)}\,.
\label{pressure}
\eq

Next introduce a new variable, the usage of which will seem somewhat ad-hoc, but whose purpose will soon become evident.  The new variable   attribute-observable pair is the following:
\bq
\beta = \frac{P}{B_z}\quad \mathrm{and}\quad \beta_0 = \frac{P_0}{B_{0z}}\,.
\label{betapBz}
\eq
Making use of the Lagrange to Euler map for the $\hat{z}$-component of the magnetic field and the pressure, respectively given by (\ref{Bz}) and (\ref{pressure}), leads to the Lagrange to Euler map of  the form
\bq
\beta = \left. \frac{\beta_0} {\mathcal{J}} \right|_{a=a({r},t)}\,
\label{beta}
\eq
which demonstrates that the above equation is similar to (\ref{rhoEu}) and (\ref{Bz}), and that  these three obey similar Eulerian equations of motion. From (\ref{betapBz}), we see that only 2 out of $\{P, B_z, \beta \}$ can be treated as independent functions. Thus, we   proceed with the following set of variables $\{\rho, M, B_z, \psi, \beta \}$, although we shall introduce $M^c$ in place of $M$ and analyze the consequences later.  We have assembled together all the requisite apparatus for  building the action principle. We proceed now to this task.  


\subsection{Constructing the gyroviscous action}

{}From Secs.~\ref{hamsPrinc} and  \ref{buildinggeneral}  we know the  kinetic energy for  a fluid element is analogous to that of a particle, and that it must obey the closure principle.  This leads to  the first term in the action,
\bq
T[q]:=\frac1{2} \int_D \!d^2a\, \rho_0(a) |\dot q|^2=
\frac1{2} \int_D \!d^2r\, \rho  |v|^2\,.
\label{Tq2D}
\eq

Now,  consider the multiple components that make up the potential energy of the Lagrangian. The first involves the internal energy of the fluid, which is  given by  the  following functional: 
\bq
U_\mathrm{{int}}[q]:=\int_D \!d^2a\, \frac{B_{0z} \beta_0}{\mathcal{J}} =
 \int_D \!d^2r\, B_ z \beta\,,
\label{Uintq2D}
\eq
which, in light of  (\ref{betapBz}),    satisfies the closure principle.

The next component of the internal energy is the magnetic field. The field energy density  is $B^2/8\pi$;   upon scaling  away the  factor of $4\pi$ we obtain  
\bqy
U_\mathrm{{mag}}[q]&:=&\frac1{2}\int_D \!d^2a\ \left(\frac{|B_{0z}|^2}{\mathcal{J}}+ \mathcal{J}   g^{kl} \a ik \,  \a jl \frac{\p \psi_0}{\p a^i} \frac{\p \psi_0}{\p a^j}\right) \nonumber  \\
&=& \frac1{2}\int_D \!d^2r\, \left(|B_{z}|^2+|\nabla \psi|^2\right)\,,  
\label{UBq2D}
\eqy
an expression that by (\ref{BEuler}) satisfies the closure principle, while physically corresponding to the magnetic energy density. 

Finally, we introduce a novel term that will be seen to account for gyroviscosity.     Since gyroviscosity is ultimately gyroscopic in nature,  this suggests  a term of the following form:
\bq
G[q]:=\int_D \!d^2a\, \Pi^{\star}\cdot \dot{q} =
 \int_D \!d^2r\, M^{\star}\cdot v \,,
 \label{Gq}
\eq
which,  unlike the other terms that were are either independent of or quadratic in $\dot{q}$,  is linear in $\dot{q}$.
It remains to determine the form of $M^{\star}$ or its corresponding attribute $\Pi^{\star}$. From the closure principle, it is evident that  $G$ should obey (\ref{Gq}), where $M^*$ is expressible purely in terms of the observables and their Eulerian derivatives.  There are an endless number of possibilities, but we shall assume that $\Pi^*$ has the simple following form:
 \bq
{\Pi^*}^i= \frac{m}{2e} \calj \ep^{ij} \a mj \, \frac{1}{\p a^m} \left(\frac{\beta_0}{\calj}\right)\,,
\label{p*}
 \eq
which is motivated in part by the knowledge that gyroviscous effects should be linear in the magnetic moment that scales as $\beta\sim P/B$.  From (\ref{Gq}), we see that
\bq
M^{\star} = \left. \frac{\Pi^*}{\calj}  \right|_{a=a({r},t)}\,,
\label{Mstar}
\eq
which can be used in conjunction with (\ref{p*}) to conclude that
\bq
M^{\star} =  \frac{m}{2e} \nabla \times \left(\beta \hat{z} \right)\,.
\label{M*}
\eq
The $m/(2e)$ prefactor of  \eqref{p*} and \eqref{M*} can be explained by defining $L^{\star}$, the intrinsic angular momentum, according to $M^{\star}=:\nabla\times L^{\star}$,  identified as the  inherent
magnetic moment of the fluid particles due to gyro-effects.    The   magnetic moment and
the angular momentum are related via the gyromagnetic ratio that is proportional to  $m/e$.  Writing the magnetization in terms of the pressure,  the magnetic moment can be identified and this leads to the factor of 2.   This also agrees with a two-dimensional version of Braginskii \cite{braginskii65}.

Now we construct our action by combining equations (\ref{Tq2D}), (\ref{Uintq2D}), (\ref{UBq2D}) and (\ref{Gq}) as follows:
\bq
S = \int_{t_1}^{t_2}\!dt\,  \left(T[q] - U_\mathrm{{int}}[q] - U_\mathrm{{mag}}[q] + G[q]\right)\,, 
\label{gyroaction}
\eq
and we are ready to explore its consequences.

We  round off this section by providing  more rationale for  the definitions of $\Pi^{\star}$ and $M^{\star}$. In the 1970s and 80s,  Newcomb \cite{newcomb72,newcomb73,newcomb83} developed a theory of incompressible gyrofluids. Later, it was shown in the 1990s \cite{AM89,AM90,AM91} that the above action gives rise to a version of the Braginskii equations \cite{braginskii65} that is a compressible generalization of  Newcomb's models (\cite{newcomb90}). 
Momentum transport by gyroviscosity arises from microscopic charged particle gyration \cite{CC53,kaufman60}, and so it is natural to think that the mass and charge of the important species  (ions for a single fluid model) would enter. Similarly, the presence of gyration is immediately suggestive, when visualized pictorially, of the presence of a curl. From (\ref{M*}), we do see that each of these properties are indeed satisfied by $M^{\star}$. 
If we were to add an additional component to the kinetic momentum $M$, such that the continuity equation remains unchanged, it is evident that the new momentum must be divergence free. In other words, it must be the curl of another quantity, which has the dimensions of angular momentum density. This provides a second reason  for $M^{\star}$ involving  a curl. Since $M$ has already been ``used'' elsewhere, this leaves $\rho$, $\psi$, $B_z$ and $\beta$ to construct this curl. Using dimensional analysis, and the presence of $m$ and $e$ (outlined above), it is seen that (\ref{M*}) can also be justified on heuristic grounds.


\subsection{The Eulerian equations and the gyromap}
\label{Eulereqns}
We begin by giving   the Eulerian  dynamical equations for the observables $\rho$, $B_z$, $\psi$ and $\beta$. These are found from the expressions (\ref{rhoEu}), (\ref{Bz}), (\ref{Psi}) and (\ref{beta}),  respectively. 
\bqy
\frac{\p \rho}{\p t}&=&-\partial_{s}M_{s}\,,
\label{rhoevol}
\\
\frac{\p B_{z}}{\p t}&=&-\partial_{s}\left(\frac{B_{z}M_{s}}{\rho}\right)\,,
\label{Bzevol}
\\
\frac{\partial\psi}{\partial t}&=&-\frac{M_s}{\rho} \partial_s \psi\,,
\label{Psievol}
\\
\frac{\p \beta}{\p t}&=&-\partial_{s}\left(\frac{\beta M_{s}}{\rho}\right)\,.
\label{betaevol}
\eqy
The final Eulerian equation, which governs the evolution of momentum, is found  from $\delta S =0$. The computation is somewhat long and tedious, but straightforward.  Hence,  we first  present the equation and then discuss the origin of various terms.  The   Eulerian momentum equation is 
\bqy
\dot{M}_{s}&=&-\partial_{l}\left({M_{s}M_{l}}/{\rho}\right)-\partial_{s} \left(P+ {|B|^2}/{2}\right) \nonumber \\
&{\ }& \hspace{1 cm}  +B_l \partial_{l} B_s - \partial_{l} \pi_{ls}\,, 
\label{momevol}
\eqy
where the pressure  is given by (\ref{betapBz}) and  the {\it gyroviscous tensor} $\pi_{ls}$ is
\bqy
\pi_{ls}&=& N_{sjlk}\beta\partial_{k}\left(\frac{M_{j}}{\rho}\right)\, \nonumber \\
N_{sjlk}&=& \frac{m}{2e}\left(\delta_{sk}\epsilon_{jl}-\delta_{jl}\epsilon_{sk}\right)\,. 
\label{Bragtensor}
\eqy

Now consider the gyroviscous action given by (\ref{gyroaction}). On varying the kinetic energy functional we obtain $\rho_0 \ddot{q}$, which yields the terms on either side of the equality sign in (\ref{momevol}). The second term in the action, the internal energy, gives rise to the pressure gradient as expected. Similarly, the magnetic component of the internal energy, which comprises of two terms, as seen from (\ref{UBq2D}), gives rise to the magnetic pressure and the penultimate term in (\ref{momevol}). Lastly, the gyroviscous part of the action gives rise to the  gyroviscous tensor, as defined in (\ref{Bragtensor}), and constitutes the last term in (\ref{momevol}).  Note, that this gyroviscous tensor is consistent with that    of Braginskii \cite{braginskii65},  when the dissipative terms are neglected and restricted to two dimensions.  It also corresponds to the model used in \cite{MCT84} and  it is a compressible generalization of that obtained by Newcomb (\cite{newcomb90}). 

The action (\ref{gyroaction}) has two different terms that involve $\dot{q}$, and hence the canonical momentum will not be the same as $\rho \dot{q}$. In fact, defining  $L=\int d^3a\, \call$, we find that
\bq
\Pi = \frac{\de L}{\de \dot{q}} = \rho \dot{q} + \Pi^{\star}\,.
\label{gyroLagPi}
\eq
Dividing throughout by $\calj$ and evaluating the expression at $a=a(r,t)$, the Eulerian counterpart is obtained through the use of (\ref{Mkin}), (\ref{Mcan}), (\ref{Mstar}) and (\ref{M*}), 
\bq
M^c = M + M^{\star} = M + \frac{m}{2e} \nabla \times \left(\beta \hat{z} \right)\,.
\label{mc}
\eq
As we are dealing with  a two-dimensional momentum vector, we can write the above equation as 
\bq
M^c_s = M_s - \frac{m}{2e} \epsilon_{ls} \partial_{l} \beta\,, 
\label{gyromaprel}
\eq
which is  the gyromap. 

We shall return to the gyromap when discussing the Hamiltonian formulation of this gyroviscous model, and show how one can move back and forth between the two variables $M^c$ and $M$, and the ensuing consequences. 


\section{The Hamiltonian description}
\label{HamDescgen}

Hitherto, we have focused almost exclusively on the action principle formulation. But, there exists a close and sometimes  bijective relationship between the action and Hamiltonian formulations; for this reason we have  grouped  them under the same heading, HAP.   When a  Lagrangian is convex,  this relationship follows straightforwardly through the  Legendre transform.  In this section, we first  review the Legendre transform for infinite systems, thereby obtaining  the Lagrangian variable Hamiltonian description.  Then, we transform to obtain  an Eulerian variable Hamiltonian description with a noncanonical Poisson bracket.   


\subsection{Legendre transformation to  Hamiltonian Form}

The Legendre transform for  infinite degree-of-freedom systems proceeds as for the finite counterpart (e.g., \cite{morrison98}).  As discussed in the preceding sections, the Lagrangian has the  form $ L=T[q] -V[q]$,  where the canonical momentum (density) is defined through  $\Pi= {\de L}/{\de \dot q}$, which for the present case is given by (\ref{gyroLagPi}).    Analogous to finite dimensions, the Hamiltonian functional is given by 
 \bq
 H[q,\Pi]=  \int_D \!d^2a\, \dot q\cdot \Pi - L\,.
 \eq
Here eliminate  $\dot q$ by expressing it in terms of $\Pi$, yielding  a Hamiltonian with a term linear in $\Pi$,  akin to that for a particle in a magnetic field.  Later we will write the Eulerian form of this Hamiltonian for our   model.

 The Poisson bracket of two functionals ${F,G}$, again invoking analogy with finite dimensions, is   given by
 \bq
 \{F,G\}= \int_D \!d^2a\,\left( 
 \frac{\de F}{\de q}\cdot  \frac{\de G}{\de\Pi}
 -  \frac{\de G}{\de q}\cdot  \frac{\de F}{\de \Pi}
 \right)\,.
 \label{canBKT}
 \eq 
 This bracket with $H[q,\Pi]$ give   equations equivalent to those of  $\de S=0$  as 
 \bq
 \dot{q}=\{q,H\}\qquad \mathrm{and}\qquad  \dot{\Pi}=\{\Pi,H\}\,,
 \eq
 the Lagrangian variable Hamiltonian form. 


 \subsection{Noncanonical Poisson Brackets and Casimirs}
 
Next we obtain from the Lagrangian variable Hamiltonian form,  an  Eulerian variable Hamiltonian form.  Because Eulerian variables are not canonical,  the Poisson bracket obtained is  noncanonical.  The idea that common models such as MHD and hydrodynamics are noncanonical Hamiltonian theories, i.e. that the theory is expressed in terms of noncanonical variables and bracket, was introduced   in \cite{morrison80}. The noncanonical bracket has Lie algebraic properties, but it possesses degeneracy and is very different from (\ref{canBKT}). The presence of degeneracy gives rise to invariants known as the Casimir invariants. We will return to the Casimirs shortly; for now, we will proceed to describe the general methodology by which a noncanonical bracket can be constructed from a canonical one.

Suppose the  functionals $F$and $G$ that enter the Poisson bracket of (\ref{canBKT}) are functions of the canonical variables $(q,\Pi)$ through the Eulerian variables, i.e., $F[q,\Pi]=\bar F[\rh, \dots]$.  
Then, variation of $F$  gives 
 \bqy
 \de F&=&\int_D \!d^2a\,\left(
 \frac{\de F}{\de q}\cdot \de q  +  \frac{\de F}{\de\Pi}\cdot \de \Pi\right) =\de \bar F
 \nonumber\\
 &=&
 \int_D \!d^2r\,\left( \frac{\de \bar F}{\de \rh} \de \rh 
 + \dots \right) \,.
\label{chain}
\eqy
But, each of these Eulerian variables depends on $q$ and $\Pi$ through the  Lagrange to Euler map. Hence, we may substitute in the Eulerian variations induced by Lagrangian to obtain  expressions for  the functional derivatives with respect to $q$ and $\Pi$ in terms of the observables.   For example, with (\ref{rhoEu}) the density variation induced by $\de q$ is
 \bq
 \de \rh=-\int_D \!d^2a\,\rh_0 \nabla \de (r-q)\cdot \de q\,,
 \eq
which gives upon substitution into  (\ref{chain})  
 \bqy
 &&\int_D \!d^2a\,\left(
 \frac{\de F}{\de q}\cdot \de q  +  \frac{\de F}{\de \Pi}\cdot \de \Pi\right) \nonumber \\
 &=&  -\int_D \!d^2r\,\frac{\de \bar F}{\de \rh} \int_D \!d^2a \, \rh_0
  \nabla \de (r-q)\cdot \de q + \dots  \,.
 \eqy
Interchanging  the order of integration  and  equating    
coefficients of $\de q$  yields an expression of the form 
\bq
 \frac{\de F}{\de q} = \calo_{\rh} \frac{\de \bar F}{\de \rh}  
 + \dots \,,
 \label{chain2}
 \eq
where the $\calo$'s appearing on the RHS are operators that involve the Dirac delta functions and  integrals over $d^2r$. A similar procedure can be carried out to obtain the functional derivative with respect to $\Pi$. Once the functional derivatives with respect to $q$ and $\Pi$ are known, we can substitute the relevant expressions into (\ref{canBKT}) and obtain the   noncanonical bracket in terms of the noncanonical Eulerian variables (e.g., \cite{morrison98,morrison05}). 

Once the noncanonical bracket is obtained, the equations of motion follow from $\dot{\phi} = \{\phi, H\}$,  where $\phi$ is any  Eulerian observable and $H$ is the Hamiltonian in terms of the observables, the existence which follows from the closure principle.  By this method of derivation,  the noncanonical bracket obtained  must  satisfy  the properties of bilinearity, antisymmetry and the Jacobi identity. 

Because the Lagrange to Euler map is not one-to-one,  noncanonical brackets are degenerate, a consequence of which are     Casimir invariants (Casimirs).  We shall not delve too deeply into the theory of Casimirs,  instead, we  refer the reader to \cite{morrison98,morrison05} for general discussion,  \cite{amp0,amp1,amp2a} for application to  MHD, and  \cite{morrison87,YMD14,YM14} for subtleties about their incompleteness.  For our purposes, it suffices to state that   Casimirs can be found from $\{F, C\} = 0$ $\forall \, F$, and to investigate   their  role in studying equilibria and stability.

Casimirs give rise to variational principles for Eulerian equilibria of the form 
\bq
\de F:=\de \left({H + C}\right) = 0\,,
\label{Equil}
\eq
where $C$ represents  any number of known Casimirs.   Given such equilibria,  the energy-Casimir method  is a means for obtaining  sufficient conditions for stability.  This method  originated in the plasma literature in  \cite{KO58}, but  hearkens back to Dirichlet's work in the 19th century \cite{dirichlet} on stability of finite-dimensional Hamiltonian systems.  From (\ref{Equil}), by second variation we obtain the symmetric matrix operator 
\bq
F_{ab}:= \frac{\de^2 F}{\de \phi^a \de \phi^b}\,,
\label{Dirichletcond}
\eq
where the $\phi$'s are the Eulerian fields.   The  energy-Casimir method states  that  positive-definiteness of this operator is   a  sufficient  condition for stability.   Thus,   Casimirs through their presence in $F$, play a crucial role in this method for determining the stability, which we investigate for a simple  gyroviscous case below, but reserve comprehensive stability analyses for  future work.


\section{The gyroviscous model - Poisson bracket,  Casimirs, and reduction}
\label{Casimirs}

Now we study two different cases of the gyroviscous model and action developed in Sec.~\ref{building}.  First  we set the attribute $\psi_0$ set to zero, and consequently  $\psi$ as well, giving a  simplified model that   is more tractable yet   possesses many similarities to the full model. This facilitates observations about the  general  nature of the Hamiltonian and the bracket, and their connection to the gyromap.  In the second case, the full bracket is studied, and  corresponding bracket and the Casimirs are obtained.


\subsection{The $\psi\equiv 0$ model}

For this reduced model we  choose to work with $M^c$ of (\ref{mc}),  because it is directly obtained via the canonical momentum $\Pi$ through (\ref{Mcan}).  Thus, since $\psi \equiv 0$,  only the $\hat{z}$-component of the magnetic field is present.   From the  Legendre transformation and the closure principle,  the following Hamiltonian is obtained:   
\begin{equation}
H = \int\!d^2r\, \left( \frac{1}{2\rho}\left|M^c - \frac{m}{2e} \nabla \times \left(\beta \hat{z} \right)\right|^2 + \beta B_z + \frac{B_z^2}{2} \right) \,, 
\label{gyroH1}
\end{equation}
which,  when written in terms of the 'kinetic' momentum $M$ becomes  the recognizable
 \bq
H = \int\!d^2r\,  \left(\frac{|M|^2}{2\rho} + \beta B_z + \frac{B_z^2}{2} \right) \,.
\label{gyroH2}
\eq
In terms of  $M^c$,  the noncanonical bracket obtained by using the procedure of  Sec.~\ref{HamDescgen}, is  
\bqy
\{F,G\}^0_c&=&\int \! d^2r \Bigg[M^c_{l}\left(\frac{\delta G}{\delta M^c_{k}}\partial_{k}\frac{\delta F}{\delta M^c_{l}}-\frac{\delta F}{\delta M^c_{k}}\partial_{k}\frac{\delta G}{\delta M^c_{l}}\right) \nonumber \\
&\quad\quad&+\rho\left(\frac{\delta G}{\delta M^c_{k}}\partial_{k}\frac{\delta F}{\delta\rho}-\frac{\delta F}{\delta M^c_{k}}\partial_{k}\frac{\delta G}{\delta\rho}\right) \nonumber \\
&\quad\quad&+B_{z}\left(\frac{\delta G}{\delta M^c_{k}}\partial_{k}\frac{\delta F}{\delta B_{z}}-\frac{\delta F}{\delta M^c_{k}}\partial_{k}\frac{\delta G}{\delta B_{z}}\right) \nonumber \\
&\quad\quad&+\beta\left(\frac{\delta G}{\delta M^c_{k}}\partial_{k}\frac{\delta F}{\delta\beta}-\frac{\delta F}{\delta M^c_{k}}\partial_{k}\frac{\delta G}{\delta\beta}\right)\Bigg]\,,  
\label{gyrobrack1}
\eqy
 the bracket for MHD of \cite{morrison82} restricted to $B=B_z\hat{z}$  with the momentum $M^c$ replacing $M$.   Thus,  the effect of  gyroviscosity is containined in  the definition of  $M^c$.   However, if we write the bracket in terms of $M$,  it becomes
\bq
\{F,G\}_G^0 =\{F,G\}^0 -\beta N_{ijsl}
\left(\partial_{s}\frac{\delta G}{\delta M_{i}}\right)\left(\partial_{l}\frac{\delta F}{\delta M_{j}}\right)\,, 
\label{gyrobrack2}
\eq
where we obtain a new term that  produces the  gyroviscous tensor and  $\{F,G\}^0$ is the bracket of (\ref{gyrobrack1}) with $M_c$ replaced by $M$. This bracket is identical to that given in \cite{MCT84}, which was obtained by ad hoc means.   If we compare the Hamiltonian-bracket pair of Eqs.~(\ref{gyroH1}) and (\ref{gyrobrack1}) with  that of  Eqs.~(\ref{gyroH2}) and (\ref{gyrobrack2}), the significance of the gyromap becomes evident. We can  choose to work with a system that possesses a relatively simple Hamiltonian with a  more complex bracket,  or vice versa, and it is the gyromap that allows us to move back and forth between these two versions. Both versions  give the same equations of motion:  those obtained in Sec.~\ref{Eulereqns} with  $\psi\equiv 0$.

Using $\{F,C\}=0$ for all $F$  to find the Casimirs implies  that the only Casimirs that exist are independent of the gyro term, and in fact, are independent of the velocity of the fluid.  We find the following infinite family of Casimirs: 
\begin{equation}
C=\int\!d^2r \, \beta f\left(\frac{\rho}{\beta},\frac{B_{z}}{\beta}\right)\,,
\label{Cas1}
\end{equation}
where  $f$ is an arbitrary function, a result that was first obtained in  \cite{MCT84}.   Because of its homogeneous form,  the three variables of (\ref{Cas1})  are interchangeable, i.e.,  we can permute $\rho$, $\beta$ and $B_z$. 

Effecting the energy-Casimir method,  as described in Sec.~\ref{HamDescgen}, we examine  equilibria  satisfying  $\de F = 0$. This yields  two  familiar conditions, 
\bq
M = 0 \qquad \mathrm{and} \qquad P + \frac{B_z^2}{2} = \mathrm{const} \,, 
\label{eqcond1}
\eq
conditions implying zero equilibrium flow and total pressure balance.   That these are equilibria  is easily checked directly from the equations of motion, as expected. 

In the above calculations, we have reverted back to $M$. This can be done in general through the following procedure:  find the Casimirs by working in terms of $M^c$, and then apply the gyromap to express them in terms of $M$.  Since the Hamiltonian is much simpler in terms of $M$, we can proceed to calculate  $\de F$, wherein $M$ is the variable of choice, and {\it not} $M^c$. We shall use  this  procedure in the following sections as well to express our final results in terms of $M$.

Having determined the equilibria, we   proceed to the stability analysis. Using the energy-Casimir methodology and computing the elements of (\ref{Dirichletcond}), we find that  equilibria  satisfying  (\ref{eqcond1}) are always stable, regardless of the functional form of the Casimir.  This is, of course,  to be expected and serves as a sanity check. 


\subsection{The $\psi  \not \equiv 0 $ model}
\label{ssec:psi_not}

Now consider the full model with $\psi  \not \equiv 0 $  and  magnetic field  given by (\ref{BEuler}). Since we have already highlighted the significance of the gyromap, we  proceed to  the Hamiltonian and  bracket in terms of $M^c$, noting that the simplicity of the latter is obtained at the expense of the former.
The Hamiltonian is
\bqy
H &=& \int \!d^2r\ \left( \frac{1}{2\rho}\left|M^c - \frac{m}{2e} \nabla \times \left(\beta \hat{z} \right)\right|^2 + \beta B_z
\right.
\nonumber\\
&&\hspace{2.5 cm} \left.+ \frac{B_z^2}{2} + \frac{|\nabla \psi|^2}{2} \right) \,,  
\label{gyroH1P}
\eqy
which is equal to  (\ref{gyroH1}) plus  the perpendicular magnetic energy,  and the Poisson bracket is
\bqy
\{F,G\}_c^{\psi}&=& \{F,G\}^{0}_c
\label{gyrobrack1P}
\\ 
&+&\int\!d^2r\, \nabla \psi \cdot
\left(
\frac{\delta F}{\delta M^c} \frac{\delta G}{\delta\psi}-\frac{\delta G}{\delta M^c} \frac{\delta F}{\delta\psi}\right) \,. 
\nonumber
\eqy
 Bracket (\ref{gyrobrack1P}) with Hamiltonian (\ref{gyroH1P}),  produce  the equations of motion derived in Sec.~\ref{Eulereqns}.

The presence of a $\psi$ leads to significant changes in the Casimirs obtained. Unlike the $\psi\equiv0$ case, we  obtain Casimirs that depend on $M^c$, which implies that they depend on the gyroviscous term. There are two different Casimir families. The first,  independent of $M^c$,  has the form
\begin{equation}
C=\int\!d^2r\, \mathcal{C}\left(\rho,\beta,B_{z}\right)\mathcal{K}(\psi)\,,
\label{Cas2}
\end{equation}
where ${\mathcal{C}}= \beta f\left({\rho}/{\beta},{B_{z}}/{\beta}\right)$ or an equivalent function involving a permutation of $\rho$, $\beta$ and $B_z$. The similarities with (\ref{Cas1}) are self-evident, since the two expressions only differ by $\calk(\psi)$. Setting $\calk=\mathrm{const}$ is tantamount to eliminating $\psi$ from (\ref{Cas2}), which makes it identical to (\ref{Cas1}). But, this  elimination of $\psi$ is exactly what differentiates the two models, which explains why (\ref{Cas2}) can be interpreted as an extension of (\ref{Cas1}).

Now we seek the  second   Casimir family  that depends  on $M^c$.   From   $\left\{ F,C\right\} =0$ we obtain 
\bqy
&&
\partial_l \left(M^c_k \frac{\delta C}{\delta M^c_l} \right) + M^c_l \partial_k \left( \frac{\delta C}{\delta M^c_l} \right)
 \label{homo}\\
&& \hspace{.9 cm} + \  
\rho\partial_{k}\left(\frac{\delta C}{\delta\rho}\right)+B_{z}\partial_{k}\left(\frac{\delta C}{\delta B_{z}}\right)
\nonumber\\
&&\hspace{1.75 cm}
+\beta\partial_{k}\left(\frac{\delta C}{\delta\beta}\right)-\frac{\delta C}{\delta\psi}\partial_{k}\psi=0\,,
\nonumber
\\
&&\partial_{k}\left(\frac{\delta C}{\delta M_{k}^{c}}\rho\right)=0\,, \qquad
\partial_{k}\left(\frac{\delta C}{\delta M_{k}^{c}}\beta\right)=0\,,
\label{mrho}
\\
&&\partial_{k}\left(\frac{\delta C}{\delta M_{k}^{c}}B_{z}\right)=0\,,
\qquad
\frac{\delta C}{\delta M_{k}^{c}}\partial_{k}\psi=0\,,
\label{psicon}
\eqy
{}From the equation of \eqref{psicon} we obtain the candidate
\bq
C=\int \! d^{2}r\, M^{c}\cdot\left(\hat{z}\times\nabla\psi\right)F\left(\rho,\beta,B_{z},\psi\right)\,, 
\eq
which when inserted in the first equation of \eqref{mrho} gives
\bq
C=\int\! d^{2}r\, \frac{M^{c}\cdot\left(\hat{z}\times\nabla\psi\right)}{\rho}\mathcal{F}\left(\psi\right)\,,
\label{C2}
\eq
while the remaining two equations of  \eqref{mrho} and  \eqref{psicon} imply
\bq
\left[\psi,\frac{B_{z}}{\rho}\right]=\left[\psi,\frac{\beta}{\rho}\right]=0,
\label{overlab}
\eq
where in cartesian coordinates $\left[f,g\right]=f_{x}g_{y}-f_{y}g_{x}$.  Equation  \eqref{overlab} implies there are no velocity dependent Casimirs unless the model is reduced, which is well known for MHD (e.g.\, \cite{amp1}).   The constraints of \eqref{overlab} are a consequence of over labeling  \cite{TM98}, since the three advected labels of Eqs.~\eqref{Bzevol}, \eqref{Psievol}, and \eqref{betaevol} cannot be independent.   Thus, we assume $B_z/\rho$ and $\be/\rho$ are functions of $\psi$ and eliminated them from the dynamics.   With this assumption \eqref{C2} is a Casimir since it also satisfies \eqref{homo}.  Upon collapsing  \eqref{Cas2},  our general Casimir is then 
\bq
C=\int\! d^{2}r\,\left( \frac{M^{c}\cdot\left(\hat{z}\times\nabla\psi\right)}{\rho}\mathcal{F}\left(\psi\right)
+
\rho\calj(\psi)
\right)\,.
\label{gencas}
\eq
Using  $B^{\perp} = \hat{z}\times\nabla\psi$,  $M \cdot B = M \cdot B^{\perp}$ and   $M^c \cdot \hat{z}=0$  (although  parallel momentum  could be included), and setting $\calf=$ constant,     the first term of \eqref{gencas}  reduces to the well-known cross-helicity invariant, except  the velocity is now $v^c = M^c /\rho$.  Thus, in  the absence of gyroviscosity,  $v^c=v$ and the usual cross-helicity is recovered.

Given the Casimir invariants we can proceed to  the variational equilibrium analysis and follow the development  of \cite{amp1}, in order  to address  the effect of gyroviscosity.   Because this  analysis  can be   complicated,   consider first the case with no flow  as a warm-up.  For  this case, the variational principle  $\de F = 0$ contains only the Casimir of   \eqref{Cas2}, giving  
\bq
M \equiv 0 \quad \mathrm{and} \quad
\Delta\psi=- P'- B_{z} B'_{z}\,, 
\eq
where $P$ and $B_{z}$ are flux functions and prime denotes differentiation with respect to $\psi$.   As expected, we obtain the  Grad-Shafranov equation.

Next consider the case with the Casimir of  \eqref{gencas}.  Since this requires the reduction due to    \eqref{overlab},  we introduce  $B_{z}=\rho \,\varpi(\psi)$ and $\beta=\rho\, \varsigma (\psi)$ and the Hamiltonian becomes  
\bqy
H&=&\int\! d^{2}r\, \Big(\frac{1}{2\rho}\left|M^{c}-M^{\star}\right|^{2}
\label{conham}\\
&&\hspace{.5 cm} + \ \rho^{2}\left[ \varsigma\varpi+\frac{\varpi^{2}}{2}\right]+\frac{|\nabla\psi|^{2}}{2}\Big)\,. 
\nonumber
\eqy
The equilibrium conditions  that follow from $\de F=0$, with \eqref{gencas} and \eqref{conham}, are 
\bqy
\frac{\delta F}{\delta M^{c}}&=&M^{c}-M^{\star}+\left(\hat{z}\times\nabla\psi\right)\mathcal{F}=0,
\\
\frac{\delta F}{\delta\rho}&=&-\frac{1}{2\rho^{2}}\left|M^{c}-M^{\star}\right|^{2} -\frac{M^{c}\cdot\left(\hat{z}\times\nabla\psi\right)}{\rho^{2}}\mathcal{F} 
\nonumber
\\ &&\hspace{.5 cm} +\  \calj  +2\rho\left[\varsigma\varpi+\frac{\varpi^{2}}{2}\right]=0,
\\
\frac{\delta F}{\delta\psi}&=&-\Delta\psi+\rho^{2}\left[\varsigma'\varpi+\varsigma\varpi'+\varpi\varpi_{\psi}\right]
\nonumber
\\ &&\hspace{.5 cm} +\ {\cal F}\nabla\cdot\left(\frac{\hat{z}\times M^{c}}{\rho}\right)+\rho \calj'=0.
\eqy
Manipulation of the above equations gives 
\bqy
&& \frac{1}{4}\left|\nabla\psi\right|^{2}\left(\frac{{\cal F}}{\rho}\right)^{2} 
+\frac{P_{z}}{\rho}
\label{bern1}
\\
&&\hspace{1.25 cm} +\ {\cal J}+\frac{m}{2e}\frac{{\cal F}}{2\rho^{2}}\nabla\beta\cdot\nabla\psi=0\,,
\nonumber\\
&& \nabla\cdot\left[\left(1-\frac{{\cal F}^{2}}{\rho}\right)\nabla\psi\right] 
+\left|\nabla\psi\right|^{2}\frac{\calf\calf' }{\rho} 
=\rho\calj'
\nonumber\\
&&\hspace{1.25 cm}  -\  \rho^{2} \left(\frac{P_z}{\rho^{2}}\right)' 
-\frac{m}{2e}{\cal F}\nabla\cdot\left(\frac{\nabla\beta}{\rho}\right)
\label{gsf1}
\eqy
with $P_{z}:=P+ {B_{z}^{2}}/{2}=\rho^{2}\left( \varsigma \varpi+{\varpi^{2}}/{2}\right)$ and recall $\beta=\rho\, \varsigma(\psi)$.   Equations  \eqref{bern1} and \eqref{gsf1} compare with those of ordinary MHD as in \cite{amp1}, but with the addition of new gyro  terms identified by the factor of $m/(2e)$.  
As for MHD,  there are free functions of $\psi$ that can be  chosen to determine current and flow profiles.  Equation \eqref{gsf1} is a generalization of the Grad-Shafranov equation, but since the density is not a flux function it alone cannot be solved.  One uses  \eqref{bern1}, a generalized Bernoulli equation,   to close the system.    These equations  are   gyro generalizations with flow of the  JOKF equation \cite{jokf}.

There are various ways of rewriting \eqref{bern1} and \eqref{gsf1}, one that brings out the Mach singularity is the following:
\bqy
&&\left|\nabla\psi\right|^{2}\left[\frac{1}{4}\left(\frac{{\cal F}}{\rho}\right)^{2} 
+  \frac{m}{2e}\frac{{\cal F} \varsigma'}{2\rho}\right]+\frac{P_z}{\rho}+{\cal J}
\nonumber\\
&&\hspace{3 cm} +\  \frac{m}{2e}\frac{{\cal F}\varsigma}{2\rho^{2}}\nabla\rho\cdot\nabla\psi=0\,,
\label{bn2}\\
&& \nabla\cdot\left[\left(1-\frac{{\cal F}^{2}}{\rho} 
+\frac{m}{2e}{\cal F}\varsigma'\right)\nabla\psi\right]
\nonumber
\\
&&\hspace{.7 cm} +\  
   \left|\nabla\psi\right|^{2}\left(\frac{\calf\calf'}{\rho}-\frac{m}{2e}{\calf' \varsigma'}\right)+\frac{m}{2e}\frac{{\cal F}\varsigma'}{\rho}\nabla\rho\cdot\nabla\psi 
\nonumber\\
   &&\hspace{.8 cm} +\   \frac{m}{2e}{\cal F}\varsigma \nabla\cdot\left(\frac{\nabla\rho}{\rho}\right)=\rho \calj' -\rho^{2} \left(\frac{P_z}{\rho^{2}}\right)'\,.
\label{gsf2}
\eqy
Evidently the equilibrium equations of \eqref{bn2} and \eqref{gsf2}  possess  a rich structure.  Analyses of their region of hyperbolicity,  modification of the fast and slow magnetosonic waves due to the gyroviscous terms,  etc. area beyond the scope of the present  paper.


\subsection{High-$\beta$  gyro-RMHD}

As noted in  Sec.~\ref{sec:intro} there exists a large literature on reduced gyrofluid models that have been obtained by various means.  Here we demonstrate how the nondissipative portion of such models can be obtained from the HAP formalism.  In particular, we show how to obtain a version of the three-field model given in  Sec.~IIIA of  \cite{HHM87}.  

From the action,  we obtained, without approximation,   the Poisson bracket of \eqref{gyrobrack1P},  or upon using the gyromap on  \eqref{gyrobrack1P} to obtain $\{F,G\}^{\psi}_G$.   If we assume  $B_{z}\rightarrow B_{0}$ and  $\rho\rightarrow \rho_{0}$ are constant, then  $P\propto\beta$,   the latter  is consistent with  incompressibility and permits us to introduce the scalar
vorticity $\Omega^{c}=\hat{z}\cdot\nabla\times M^{c}$ and  $M^{c}=\nabla\varphi^{c}\times\hat{z}$,
where $\varphi^{c}$ is the stream function, up to the constant factor  of $\rho_0$. 

The subscript $c$ is present everywhere to indicate that these include the gyroviscous terms. Following a similar line of analysis to that  employed in  \cite{amp0}, viz.\ chain rule relations of the form $\nabla^2\de F/\de\Om^c=\hat{z}\cdot \nabla\times\de F/\de M^c$, we  reduce the bracket of \eqref{gyrobrack1P} to the following:
\bqy
\left\{ F,G\right\} &=&\int\! d^{2}r\Big(\Omega^{c}\left[F_{\Omega^{c}},G_{\Omega^{c}}\right]
\label{rmhd}\\
&&\hspace{.50 cm}
+\ \psi\left(\left[F_{\psi},G_{\Omega^{c}}\right]-\left[G_{\psi},F_{\Omega^{c}}\right]\right)
 \nonumber\\
&&\hspace{1.25 cm} + \ \beta\left(\left[F_{\beta},G_{\Omega^{c}}\right]-\left[G_{\beta},F_{\Omega^{c}}\right]\right)
\Big)\,,
\nonumber
\eqy
which is precisely the high-$\beta$ RMHD bracket first given in \cite{MH84}.  Because \eqref{rmhd} is 
homogeneous of degree zero in $\beta$ and $\psi$ and of degree one in $\Om^c$, which means scaling $\Om^c$ only scales time,  these quantities can be identified with the corresponding quantities of  \cite{HHM87}.   Our Hamiltonian reduces to 
\begin{equation}
H=\frac12\int\! d^{2}r\Big({\left|\nabla\varphi\right|^{2}}
+{\left|\nabla\psi\right|^{2}}\Big)\,,
\label{betaHam}
\end{equation}
which has no pressure terms,  because for simplicity  we neglect the effect of toroidal curvature that usually occurs in high-$\be$ RMHD.

From \eqref{mc},  the  gyromap   relation between $M^{c}$ and
$M$, we obtain 
\begin{equation}
\varphi^{c}=\varphi+\frac{m}{2e}\beta\,,
\label{gmap2}
\end{equation}
where  $M=\nabla \varphi\times \hat{z}$.   Equation \eqref{gmap2} is precisely the gyromap used in \cite{HHM87}.  Using \eqref{gmap2} we can follow one of two paths:  eliminate $\varphi$ from   \eqref{betaHam} and insert the resulting  $H$ into \eqref{rmhd} to obtain the equations of motion in terms of $\Om^c$, or transform the bracket of   \eqref{rmhd}  to one in terms of $\Om$ and use the Hamiltonian of  \eqref{betaHam}  as it stands.  Both give gyrofluid evolution equations  equivalent to those  of  \cite{HHM87},  with the neglect of toroidal curvature and a Hall term in Ohm's law that is an extended MHD effect outside the scope of the present theory.


\section{Conclusion}
\label{Conclusion}

In this paper we have described a general procedure for constructing action principles for continuum models, an important portion of which is the elucidation of the Eulerian closure principle.  The procedure was used to construct a fluid model with gyroviscous effects, and it was also shown how to obtain the Eulerian Hamiltonian formalism.  Consequences of the construction are the following:  the unambiguous identification of the origin of the gyromap,  Casimir invariants for the gyroviscous fluid models,  variational principles for equilibria  with flow giving rise to generalized Grad-Shafranov equations with gyroviscous effects, and a first principles derivation of the gyroviscous effects that appear in reduced fluid models.

We believe that the tools developed here, the method of constructing action principles and concomitant Hamiltonian  formalisms for fluid models  with,  in particular,  the incorporation of  finite Larmor radius (FLR) effects,  provide  a natural, efficient,  and transparent method for deriving the nondissipative parts  of plasma theories.   The methodology is quite general --  a natural extensions of the present work would be to include additional two-fluid effects, giving rise to, e.g., the gyroviscous cancellation  \cite{HH71,RT62,HKM85}, and to obtain generalizations of  models such as that of   \cite{HHM86}.


\end{document}